\documentclass[aps,amsmath,amssymb,notitlepage]{revtex4-1}

\usepackage{graphicx}
\usepackage{amsmath}
\usepackage{hyphenat}
\usepackage{siunitx}

\DeclareMathOperator\erf{erf}

\usepackage[caption=false]{subfig}

\usepackage{color}

\begin{document}

\title{Correcting for accidental correlations in saturated avalanche photodiodes}

\author{J.A. Grieve}
\email[For correspondance: ]{james.grieve@nus.edu.sg}
\author{R. Chandrasekara}
\author{Z. Tang}
\author{C. Cheng}

\affiliation{Centre for Quantum Technologies, National University of Singapore,\\Block S15, 3 Science Drive 2, 117543 Singapore}

\author{A. Ling}

\affiliation{Centre for Quantum Technologies, National University of Singapore,\\Block S15, 3 Science Drive 2, 117543 Singapore}

\affiliation{Department of Physics, National University of Singapore,\\Block S12, 2 Science Drive 3, 117551 Singapore}

\date{\today}

\begin{abstract}
In this paper we present a general method for estimating rates of accidental coincidence between a pair of single photon detectors operated within their saturation regimes. By folding the effects of recovery time of both detectors and the detection circuit into an ``effective duty cycle'' we are able to accomodate complex recovery behaviour at high event rates. As an example, we provide a detailed high-level model for the behaviour of passively quenched avalanche photodiodes, and demonstrate effective background subtraction at rates commonly associated with detector saturation. We show that by post-processing using the updated model, we observe an improvement in polarization correlation visibility from 88.7\% to 96.9\% in our experimental dataset. This technique will be useful in improving the signal-to-noise ratio in applications which depend on coincidence measurements, especially in situations where rapid changes in flux may cause detector saturation. 
\end{abstract}

\maketitle

\section{Introduction}

Avalanche photodiodes operating in Geiger mode (GM-APDs) \cite{brown86,brown87,dautet93,zappa96} are a workhorse tool in discrete quantum optics experiments that depend on the correlated detection of separated single photons. Such experiments include fundamental tests of Bell's Inequality~\cite{shih88}, quantum communication~\cite{gisin02} and correlated photon imaging or microscopy ~\cite{malik12,schlenk14,morris15}.  Due to their small size and relatively low cost GM-APDs are enabling the deployment of quantum technologies in real-world scenarios \cite{Ling2008,tang14}.

In these experiments utilizing correlated photons, the signal is coincident events in two (or more) single photon detectors. Achievable signal rates are bounded by the maximum detection rate of each detector. In practice, the finite recovery time of the single photon detectors results in the onset of saturation behaviour far below the maximum observable rate. In this so-called ``saturation regime'', a rate dependent detection efficiency comes to dominate the individual event rates, with a knock-on effect that decreases the signal-to-noise ratio for coincidences. Due to the difficulty in estimating the noise (accidental coincidences), most laboratory-based experiments are effectively rate limited.

While such behaviour is present in all single photon detectors, the problem is especially pronounced when utilising GM-APDs in the passively quenched configuration~\cite{brown86}. While the relatively low cost, simplicity and power efficiency of these systems make for an attractive alternative to the more demanding actively quenched strategies, a typical recovery time of approximately \SI{1}{\micro\second} \cite{zappa96} limits the throughput of the system. In a typical passively quenched device saturation effects become dominant between one and two hundred thousand events per second, severely limiting the dynamic range achievable. This hinders the deployment of cost effective and low-power GM-APD modules for applications relying on coincidence between two or more detectors for the reasons outlined above. Although efforts to develop faster passive quenching circuits are ongoing~\cite{lunghi12} it is only possible to delay the onset of saturation effects.

In applications such as imaging by coincidence~\cite{karmakar12,schlenk14} the restricted dynamic range can be a severe limitation. In some very interesting situations such as coincidence imaging of a scene under sunlight illumination~\cite{karmakar12,liu14} changing flux due to environmental conditions is to be expected. In other applications such as coincidence based microscopy~\cite{schlenk14}, changing flux can arise from bright regions. The operator must then accept the loss in signal contrast or be motivated to use lengthy data acquisition times. 

Contrast could be improved and integration times shortened by permitting detectors to operate deep into their saturation regions, with data quality and signal contrast preserved by post-processing to remove accidental coincidences. In this manuscript we present a model for accidental coincidence estimation based on the concept of an ``effective duty cycle''. This method may be applied to virtually any single photon detector technology provided that the recovery process is well known. 

\section{The effective duty cycle}
In a simple coincidence measurement strategy digital pulses from two GM-APDs are sent to an electronic AND gate [12] where overlapping pulses give a coincidence event. Two independent streams of photons from uncorrelated sources will occasionally produce coincidences~\cite{Polyakov2006, Wyllie1987} termed accidental coincidences. When detectors are operating far below their saturation thresholds the rate of accidental coincidences, $C_{acc}$ can be estimated by:

    \begin{equation}
        \label{eqn:old-correction-asynch}
        C_{acc} = S_1 S_2 (\tau_1 + \tau_2).
    \end{equation}

\begin{figure}[h!]
        \centering
        \subfloat[\label{fig:aplusb}]{
            \includegraphics[width=0.4\linewidth,bb=0 0 320 240]{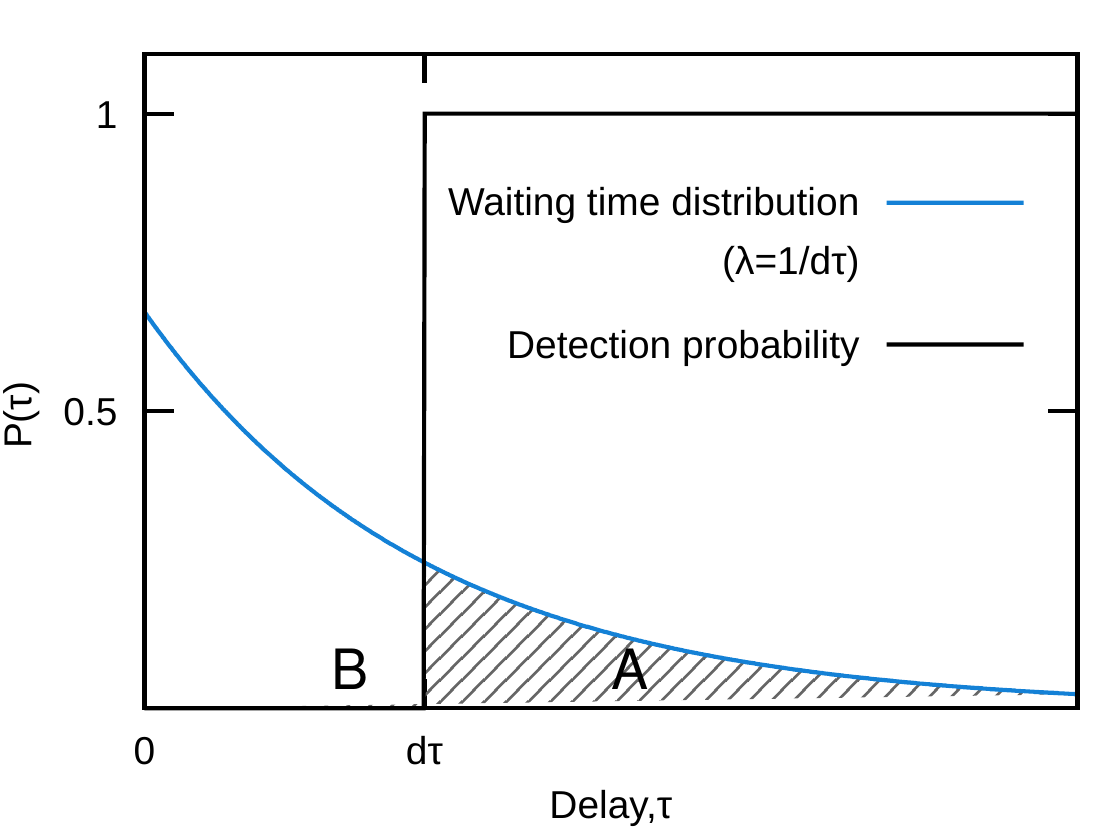}
        }
        \subfloat[\label{fig:eta-components}]{
            \includegraphics[width=0.35\linewidth,bb=-20 -50 299 162]{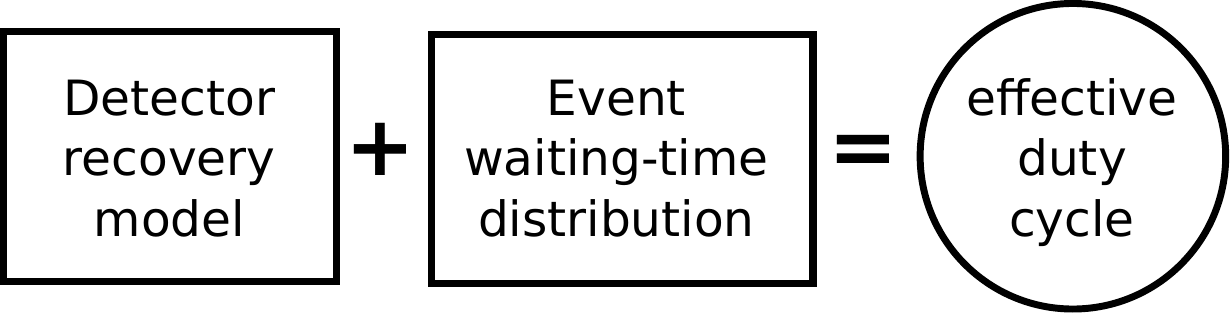}
        }
        \caption{(a) An illustration of the interplay between event rate and recovery time. The waiting time distribution for a Poissonian source (of rate $\lambda$) is shown alongside a simple stepwise model for the detection probability, exhibiting a well defined ``dead time'' ($d\tau$) with unit efficiency elsewhere. It is clear that a portion ($A$) of the event sequence will be detected, with a rate dependent efficiency, $\eta = A/(A+B)$. (b) The effective duty cycle is calculated by combining a model for the incoming event sequence, with a model for the detector recovery process.}
    \end{figure}

    Here $\tau_{1,2}$ are the duration of the digitized signal pulses sent to the AND gate and $S_{1,2}$ are the observed signal rates. Note that this model neglects the effect of finite recovery time entirely, and so is only approximately accurate even at low rates. One hypothetical recovery process is illustrated in Fig~\ref{fig:aplusb}, and represents the trivial situation of a well defined ``dead time'' immediately after an avalanche pulse with a sharp, stepwise recovery to unit efficiency. In such a system, the proportion of time a detector spends in a receptive state is rate dependent. This rate dependent receptive time can be identified as its effective duty cycle. When this is taken into account, it is clear that Eq~\ref{eqn:old-correction-asynch} leads to an underestimation because the detection events are actually taking place over a smaller fraction of time, increasing the chance of accidental overlap of digital pulses. Using similar arguments to \cite{Wyllie1987} we arrive at a corrected expression for $C_{acc}$:

    \begin{equation}
        \label{eqn:new-correction-asynch}
        C_{acc} = S_1 S_2 \left( \frac{\tau_1}{\eta_1} + \frac{\tau_2}{\eta_2} \right).
    \end{equation}

Here $\eta_{1,2}$ are the effective duty cycles of the detectors for the observed rates $S_{1,2}$. It is clear that determination of $\eta$ requires the combination of a model for the detection probability recovery process with some assumptions regarding the statistics of the light source, in particular the distribution of inter-arrival times (also known as waiting times, Fig~\ref{fig:eta-components}). 

It is also interesting to note that when Eq~\ref{eqn:new-correction-asynch} is rewritten to group contributions from each detector (note subscripts, Eq~\ref{eqn:new-correction-asynch-2}), we are also able interpret $\eta$ as the proportion of the incoming events which trigger a detection.
    
    \begin{equation}
        \label{eqn:new-correction-asynch-2}
        C_{acc} = S_2 \left( \frac{S_1\tau_1}{\eta_1} \right) + S_1\left(\frac{S_2\tau_2}{\eta_2} \right).
    \end{equation}

With this very general definition, it is clear that we need not restrict our analysis to trivial recovery processes, nor to any particular photon statistics. This approach is sufficiently general to be used with any pair of single photon detectors, and may be extended to accomodate $N$-fold coincidence measurements. In the rest of this manuscript, our discussion will be focused on obtaining the duty cycle for passively quenched GM-APDs that are operating deep into the saturation regime.

\section{A model for GM-APDs}

\begin{figure}[h!]
    \centering
    \subfloat[\label{fig:apd}]{
        \includegraphics[width=0.25\linewidth,bb=0 -60 394 302]{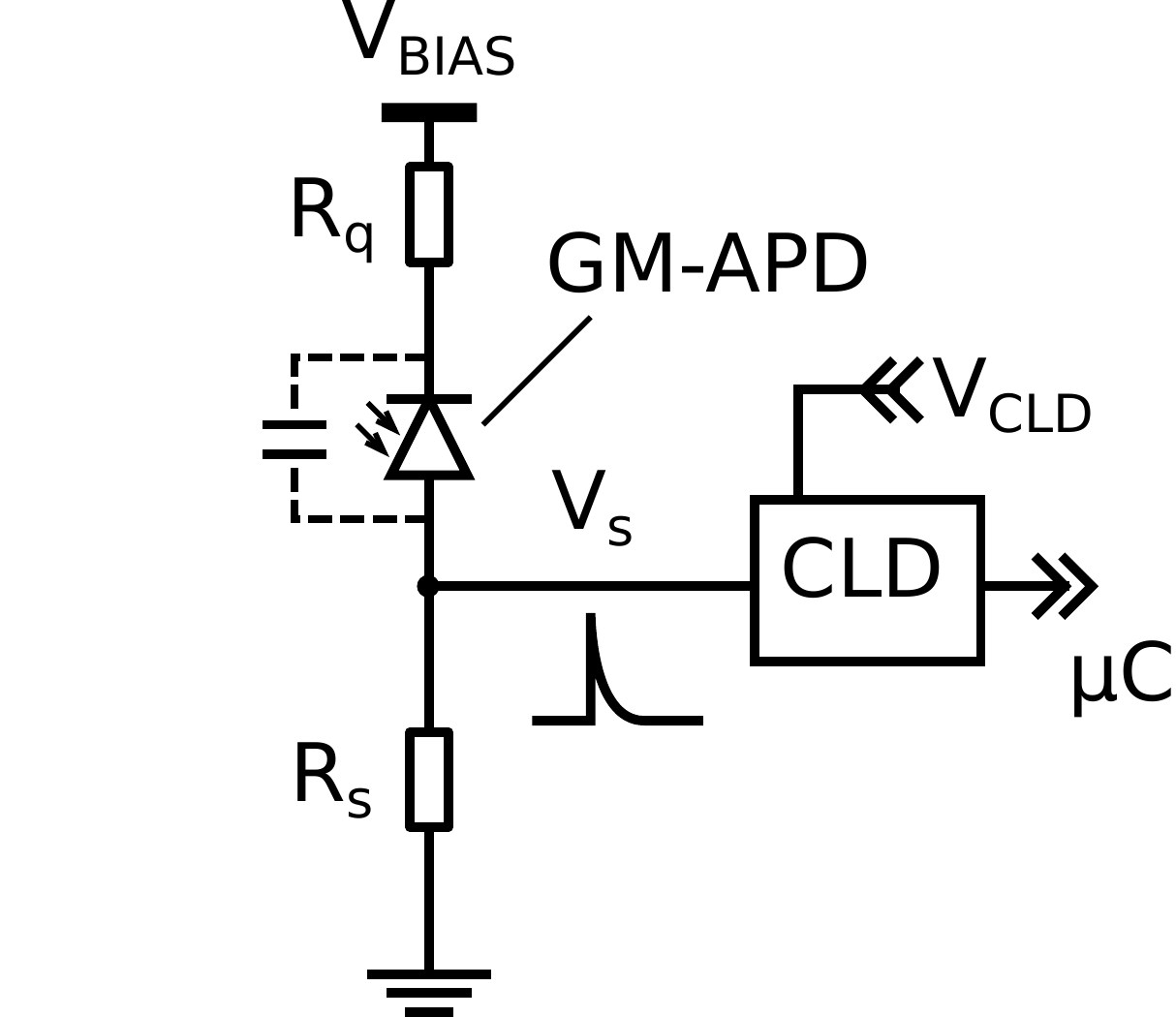}
    }
    \subfloat[\label{fig:pulse_heights}]{
        \includegraphics[width=0.4\linewidth,bb=0 0 320 240]{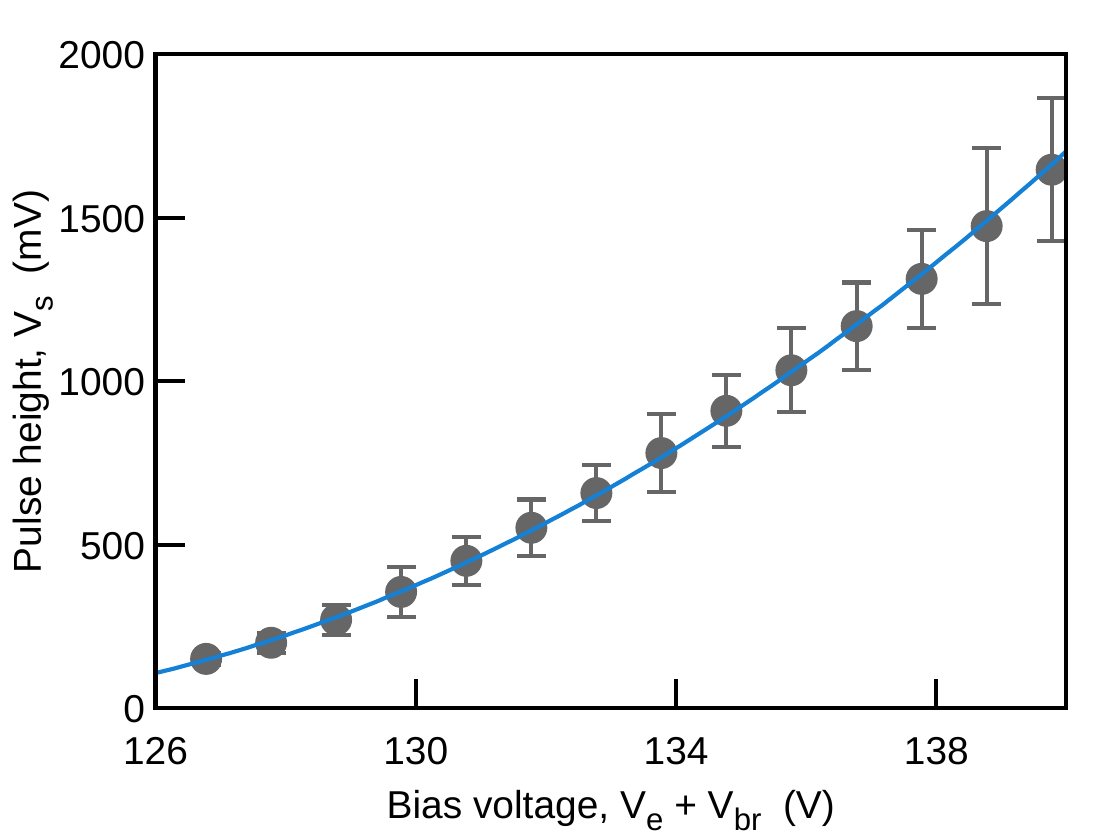}
    }
    \caption{\label{fig:apd-plus-pulse_heights} (a) A passive quenching circuit for a GM-APD where the device is in series with quench and sense resistors ($R_q$ and $R_s$). A signal pulse ($V_s$)  is monitored by a constant level discriminator ($CLD$) supplied with a reference voltage ($V_{CLD}$). This gives rise to a digital pulse, suitable for counting on a microcontroller ($\mu C$). A simple method for identifying coincidences between two GM-APDs is to observe if the electronic pulses overlap within an electronic AND gate. (b) Measured pulse heights (corresponding to $V_s$ in Eq.~\ref{eqn:Vs-vs-Ve}) in a GM-APD (SAP500 Laser Components) as a function of bias voltage. The solid line is a fit using Eq.~\ref{eqn:Vs-vs-Ve} with an offset corresponding to the breakdown voltage. The squared dependence of $V_s$ upon $V_e$ is clearly visible.}
\end{figure}

\begin{figure}[ht]
    \centering
    \includegraphics[width=0.4\linewidth, bb=0 0 320 240]{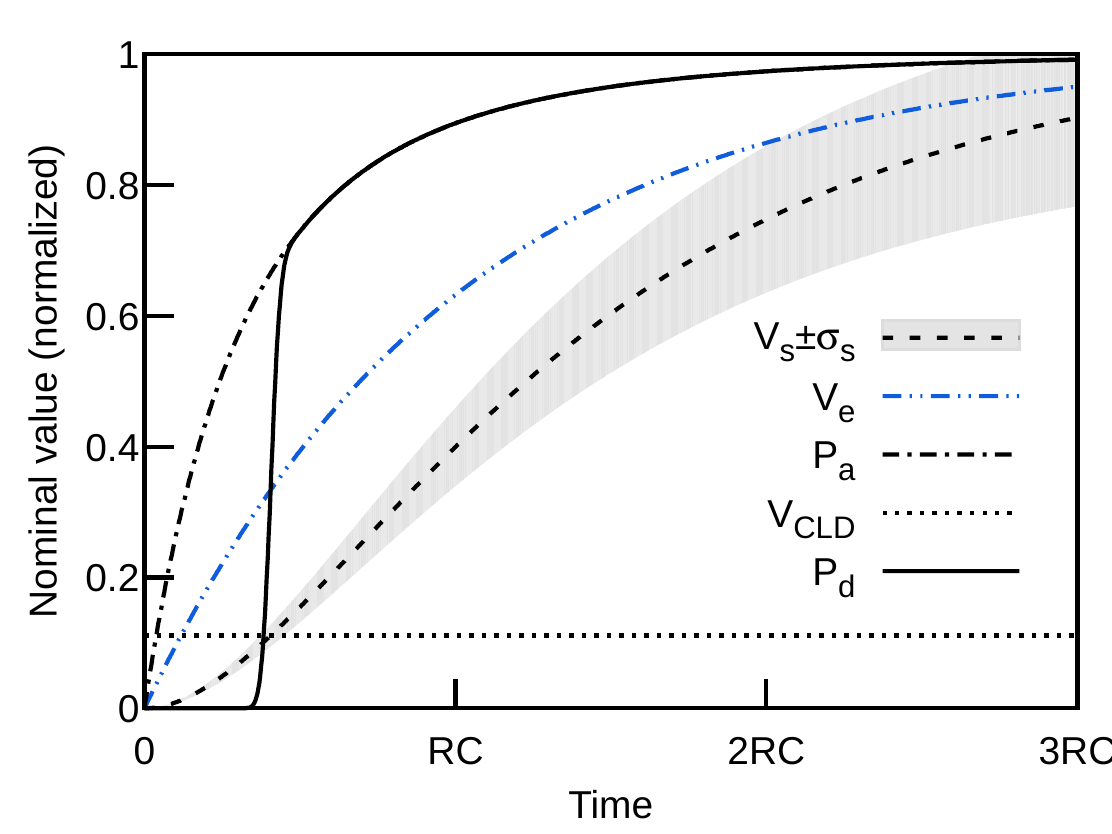}
    \caption{\label{fig:trends}Illustration of the recovery of GM-APD excess voltage, signal probability and signal voltage as a function of time after an avalanche event. By comparing these trends to the reference voltage $V_{CLD}$ (plotted, normalized against the signal voltage) we can produce a detection probability curve (solid black line). The probability is seen to be exactly zero during the period where $V_s << V_{CLD}$, and rises sharply once this value is exceeded. There is a transition point where $V_s >> V_{CLD}$ as the dominant mechanism impacting the probability becomes $P_a$, and the combined probability $P_d$ recovers towards its nominal value.}
\end{figure}

A typical passively quenched GM-APD configuration is shown in Fig.~\ref{fig:apd}. The GM-APD is reverse biased in series between two resistors, known as the quench and sense resistors $R_q$ and $R_s$. The resistor $R_q$ is selected to produce a voltage drop that will bring the detector bias below its breakdown value, halting the avalanche. The voltage at the sense resistor $R_s$ may be monitored to detect avalanche events. The behaviour of this circuit, in particular the recharging time, can be understood by considering a small number of relatively high-level relationships between just a few parameters. These are listed in Table \ref{table:symbols} and shown graphically in Fig.~\ref{fig:trends} and will be discussed below.

\begin{table}[hb]
    \centering
    \begin{tabular}{r p{0.8\linewidth}}
    \hline
    Symbol & Description \\
    \hline
    $P_a$ & Probability that a charge carrier will trigger an avalanche \\
    $P_s$ & Probability that an analogue signal is sensed by the CLD \\
    $P_d$ & Combined probability of a digital signal given a charge carrier \\
    $V_e$ & Time varying excess voltage (overvoltage) \\
    $V_s$ & Analogue signal voltage \\
    $V_{CLD}$ & Constant level discriminator reference voltage \\
    \hline
    \end{tabular}
    \caption{\label{table:symbols}A selection of symbols used throughout the text, collected for clarity. We distinguish between a number of probabilities during the avalanche and recovery process.}
\end{table}
    
    The difference between the applied bias and the GM-APD's characteristic breakdown voltage ($V_{br}$) is known as the overvoltage or excess voltage. Immediately following an avalanche this voltage falls rapidly to zero, before recovering to its nominal value. We distinguish between the excess voltage set-point ($V_{e,set}$) and the instantaneous excess voltage $V_e(t)$, given by
    
    \begin{equation}
        \label{eqn:Ve-vs-t}
        V_e(t) = V_{e,set} ( 1 - exp(-t/RC) ),
    \end{equation}
    
    \noindent where $RC$ is the characteristic recharge time dominated by the parasitic capacitance of the GM-APD (see Fig~\ref{fig:apd}) and the resistance of the quench resistor ($R_q$) and $t$ represents the elapsed time after an avalanche event. The time for the excess voltage to fall to zero may be approximated as instantaneous, after which it recovers exponentially towards its nominal set-point.
    
    This time dependence of the excess voltage gives rise to a time-varying electric field in the depletion zone. The probability that a charge carrier in the depletion zone of the GM-APD will trigger an avalanche ($P_a$) is a function of the acceleration due to this field and hence the excess voltage, and may be expressed as
    
    \begin{equation}
        \label{eqn:Pa-vs-Ve}
        P_a = 1 - exp(-V_e/V_c),
    \end{equation}
    
\noindent  where $V_c$ is the \emph{characteristic voltage} of the GM-APD and must be characterised for each device~\cite{dautet93}.
    
    Avalanche events in the GM-APDs give rise to a voltage across a sense resistor ($R_s$) with amplitude $V_s$. In our devices (SAP500, Laser Components~\cite{stipcevic13}) we observe the magnitude of this signal to depend upon the square of $V_e$ as shown in Fig.~\ref{fig:pulse_heights}:
    
    \begin{equation}
        \label{eqn:Vs-vs-Ve}
        V_s = A V_e^2.
    \end{equation}
    
     The analogue signal voltage $V_s$ must be converted into a digital pulse usually by using a constant level discriminator (CLD). The analogue signal is compared to a reference voltage ($V_{CLD}$) and a digital pulse is produced when triggered. This gives rise to a sensing probability ($P_s$) that is a binary stepwise function falling to zero when $V_e$ yields signal voltages ($V_s$) below the reference. Effects associated with the detector active area~\cite{Polyakov2006} result in a distribution of signal amplitudes ($V_s$), which we take to be approximately Gaussian. This has the effect of smoothing the transition in $P_s$, and we approximate this as
    
    \begin{equation}
        \label{eqn:Ps}
        P_s(t) = \erf((t-t_0)/\sigma_{s}),
    \end{equation}
    
\noindent where $t_0$ is the time at which the mean value of $V_s$ is equal to $V_{CLD}$ and $\sigma_s$ is the width of the distribution of $V_s$, commonly observed to be 10-20\% of $V_s$ (see Fig.~\ref{fig:pulse_heights}, error bars).  Combining this sensing probability with the avalanche probability, we arrive at our final model for detection probability,
    
     \begin{equation}
        \label{eqn:Pd}
        P_d(t) = P_s(t) P_a(V_e(t)).
    \end{equation}
    
     The curve for $P_d$ is plotted in Fig.~\ref{fig:trends} (solid black line), and clearly exhibits an ``effective dead time'' that is strongly influenced by both the choice of $V_{CLD}$ and the excess voltage setpoint $V_{e,set}$. We choose to ignore the non-unit, wavelength dependent probability of a photon giving rise to a photoelectron and restrict ourselves to the probability of detecting only charge carriers that have been created. It is assumed that photons which do not result in photoelectrons do not affect the behaviour of the detectors. The effective duty cycle concept is sufficiently broad to encompass virtually all detector behaviour, provided it is adequately described in the recovery model. For example, we draw the reader's attention to the non-zero probability of an avalanche being trigered even for very small delays (within this dead time), implying the existence of so-called ``extendable'' dead time and associated detector paralysis~\cite{schatzel86}. These effects can be clearly seen in Fig~\ref{fig:rates}, where the observed rate exhibits a maximum and subsequent reduction at high event rates.

\section{Calculating the effective duty cycle}

    In order to calculate the effective duty cycle, it is necessary to make some assumptions regarding the statistics of the incoming event sequence. When considering avalanche photodiodes, we are interested in the temporal distribution of charge carriers in the device's active area, as these are the events which trigger avalanches and subsequent digital pulses. The dominant contribution to this rate is the incoming photon flux, which we take to be Poisson distributed with a corresponding exponential distribution of waiting times determined by the rate $\lambda$. As in the previous section, we ignore the probability of a photon giving rise to a charge carrier in the detector, and make the further assmption that this sampling does not impact the statistics of the light.

    Additional contributions to the distribution of charge carriers will be present in the form of higher-order effects such as afterpulsing, twilighting~\cite{Polyakov2006} and so-called ``dark counts'' associated with thermal electrons. In passively quenched detectors operated at room temperature, the recovery process is generally slow enough to render afterpulsing and twilighting insignificant~\cite{zappa96}. Dark counts are assumed to also exhibit Poisson statistics, and therefore increase the event rate without influencing its statistics. We assume that deviations due to event bunching or anti-bunching are small enough to be insignificant, but care should be taken in making this assessment.

    \begin{figure}[t]
        \centering
        \subfloat[\label{fig:unsaturated}]{
            \includegraphics[width=0.4\linewidth,bb=0 0 320 240]{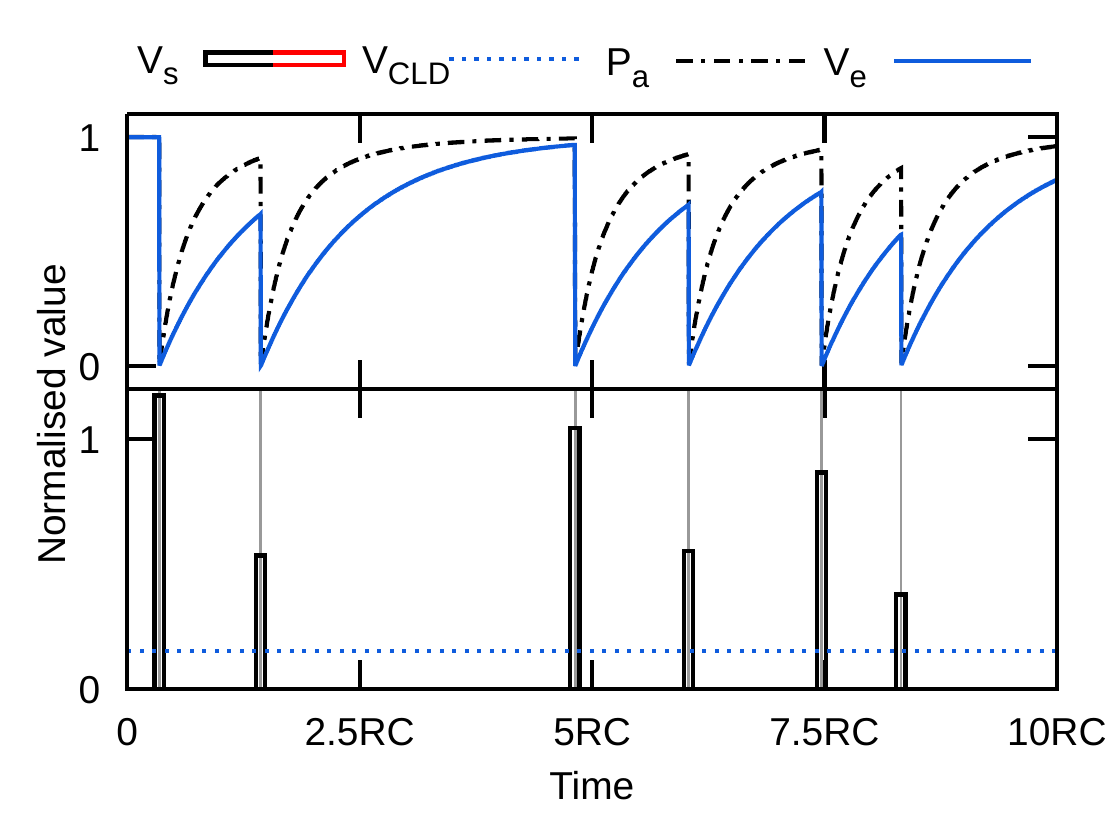}
        }
        \subfloat[\label{fig:saturated}]{
            \includegraphics[width=0.4\linewidth,bb=0 0 320 240]{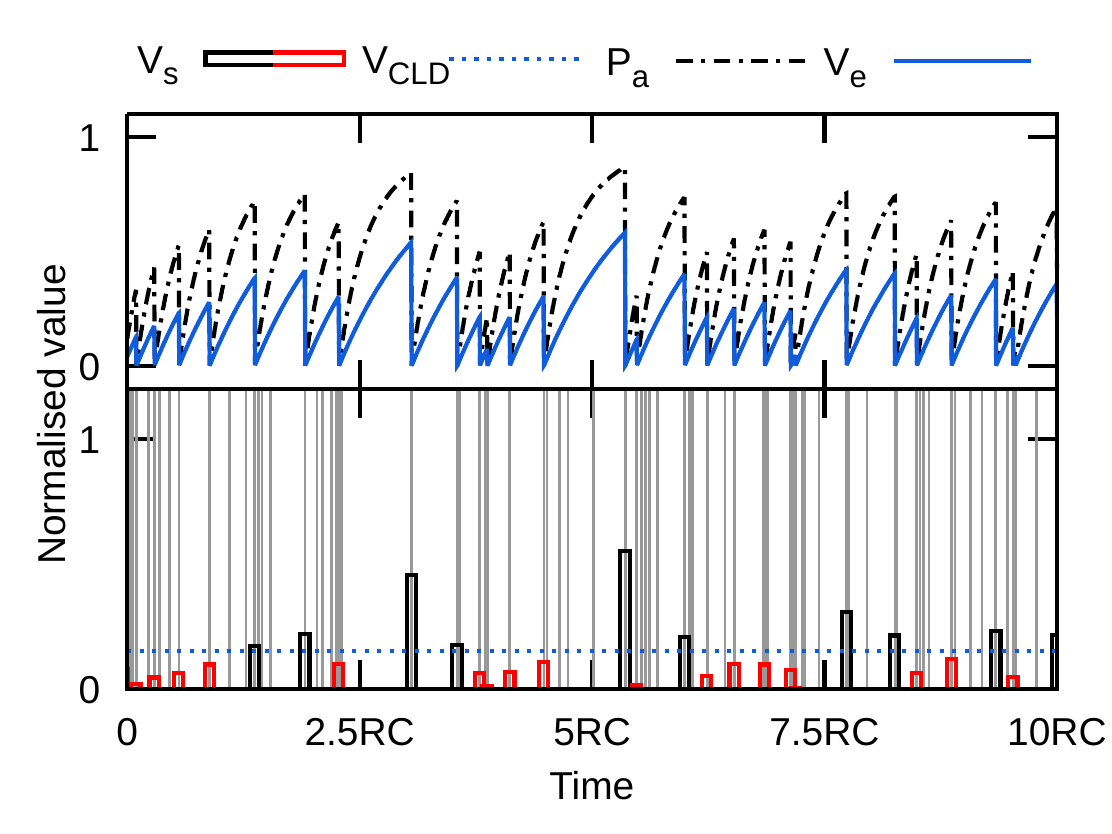}
        }
        \caption{\label{fig:satunsat} Example excess voltage ($V_e$), avalanche probability ($P_a$) and analogue pulse height ($V_s$) curves from the full numeric simulation, plotted for rates that are (a) below and (b) deep within the saturation region of the GM-APD. Also shown is the discriminator voltage ($V_{CLD}$), with analogue pulses labelled to indicate whether they are sensed (black) or ignored (red) by the circuit. Thin vertical lines indicate the position of events in the input sequence. It is clear that in the saturated case, many events do not trigger avalanches (due to the low instantaneous $P_a$ value), and of those that do, most are not sensed. The fraction of of the input sequence that goes on to produce sensed events is identified as the effective duty cycle $\eta$.  }
    \end{figure}
    
    In many cases, it may prove too challenging to develop an analytic expression for the effective duty cycle, and we found this to be the case for the passively quenched GM-APD circuits described here. Therefore, to calculate the effective duty cycle we performed a full numeric simulation of Equations~\ref{eqn:Ve-vs-t}-\ref{eqn:Pd} using the GNU Octave mathematics package. Sequences of arrial times were drawn from the exponential distribution, and the corresponding observed rates were calculated by analysis of the resulting curves. To calculate the effective duty cycle two approaches were employed. The first is to calculate the area under the $P_d$ curve and normalise for unit time, while the second is to compute the fractional rate, i.e. the ratio of observed rate to input rate. These two methods were found to agree exactly. Example data is shown in Fig~\ref{fig:satunsat} for input sequences associated with saturated and mostly unsaturated behaviour. 

\begin{figure}[t]
        \centering
        \subfloat[\label{fig:rates}]{
            \includegraphics[width=0.4\linewidth,bb=0 0 320 240]{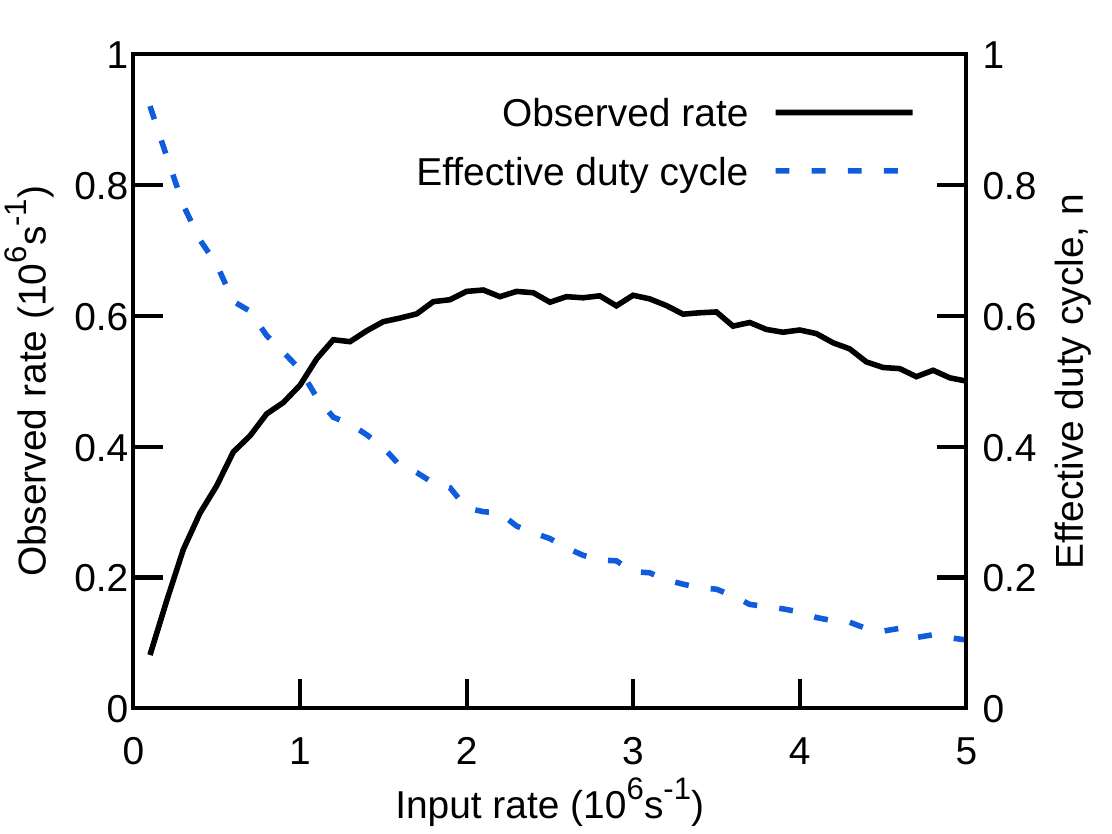}
        }
        \subfloat[\label{fig:heatmap}]{
            \includegraphics[width=0.4\linewidth,bb=0 0 320 240]{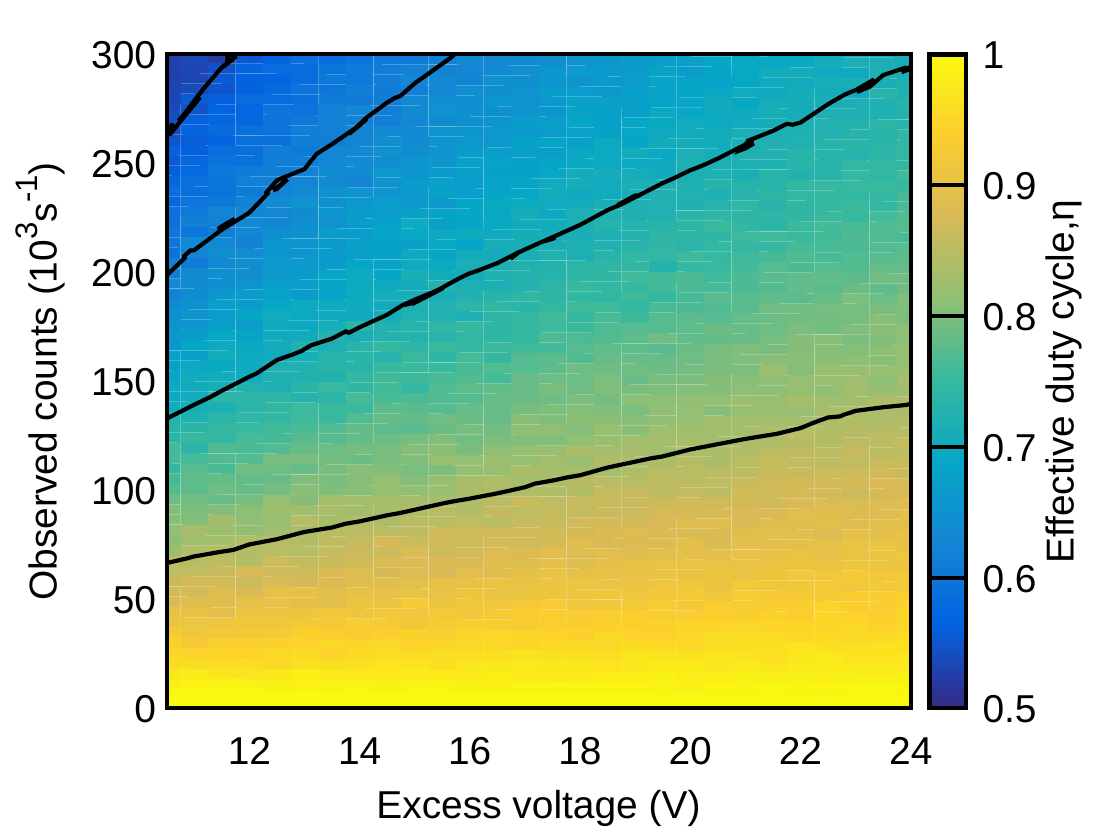}
        }
        \caption{(a) An example calculation for a GM-APD with $V_{e,nom} = 3V_c$ ($15\,V$ and $5\,V$ respectively). Values calculated from the numerically simulated curves (see Fig~\ref{fig:satunsat}) are shown against the input event rate.  (b) Calculated effective duty cycle as a function of both nominal excess voltage $V_{e,nom}$ and observed count rate. Contours are plotted at increments defined on the colorbar.}
    \end{figure}

    For use in subtracting background in an experimental setting, we calculate effective duty cycles for avalanche photodiode modules operating over a range of bias voltages and count rates, and tabulate these opposite the corresponding observed rate. Example data for an SAP500 device is plotted in Fig.~\ref{fig:heatmap}, and may be used as a lookup table when attempting to remove background counts in experiments using this particular GM-APD. We can also compare the curvature of the observed count rate and verify that this behaviour matches expectations (Fig.~\ref{fig:rates}).

\begin{figure}[t!]
    \centering
    \subfloat[\label{fig:visibility}]{
        \includegraphics[width=0.4\linewidth,bb=0 0 320 240]{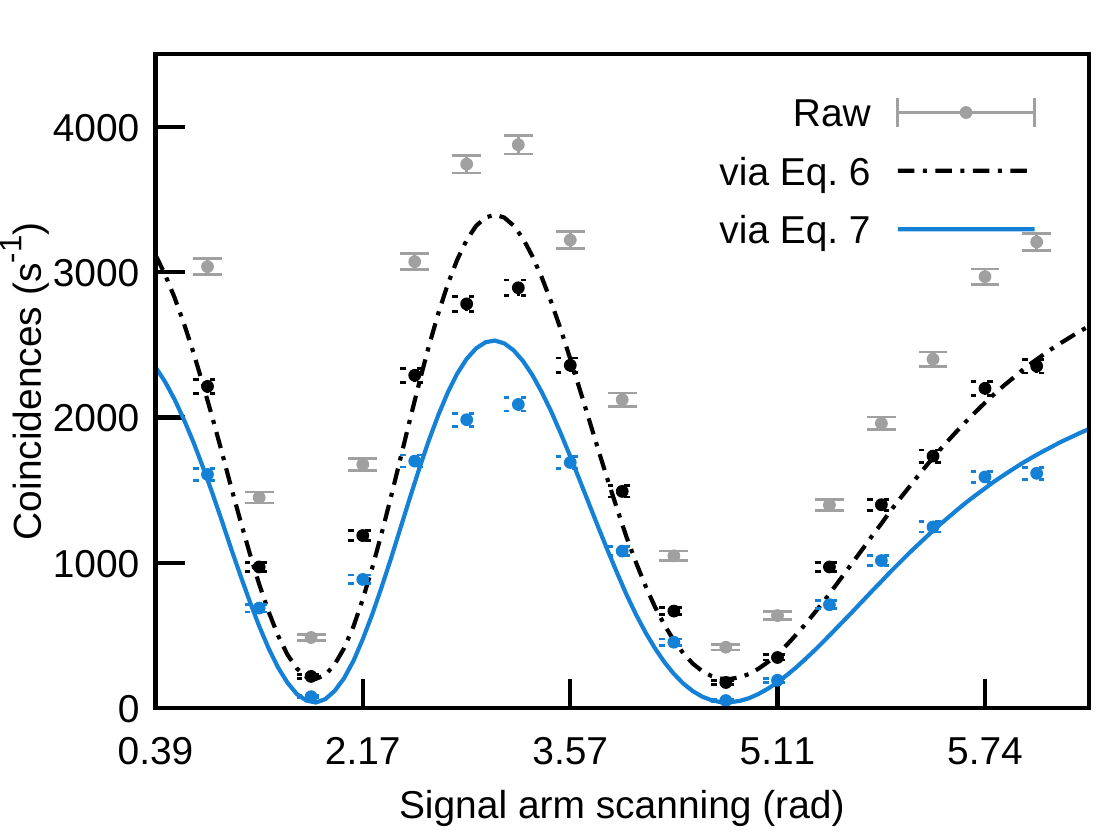}
    }
    \subfloat[\label{fig:visibilities}]{
        \includegraphics[width=0.4\linewidth,bb=0 0 320 264]{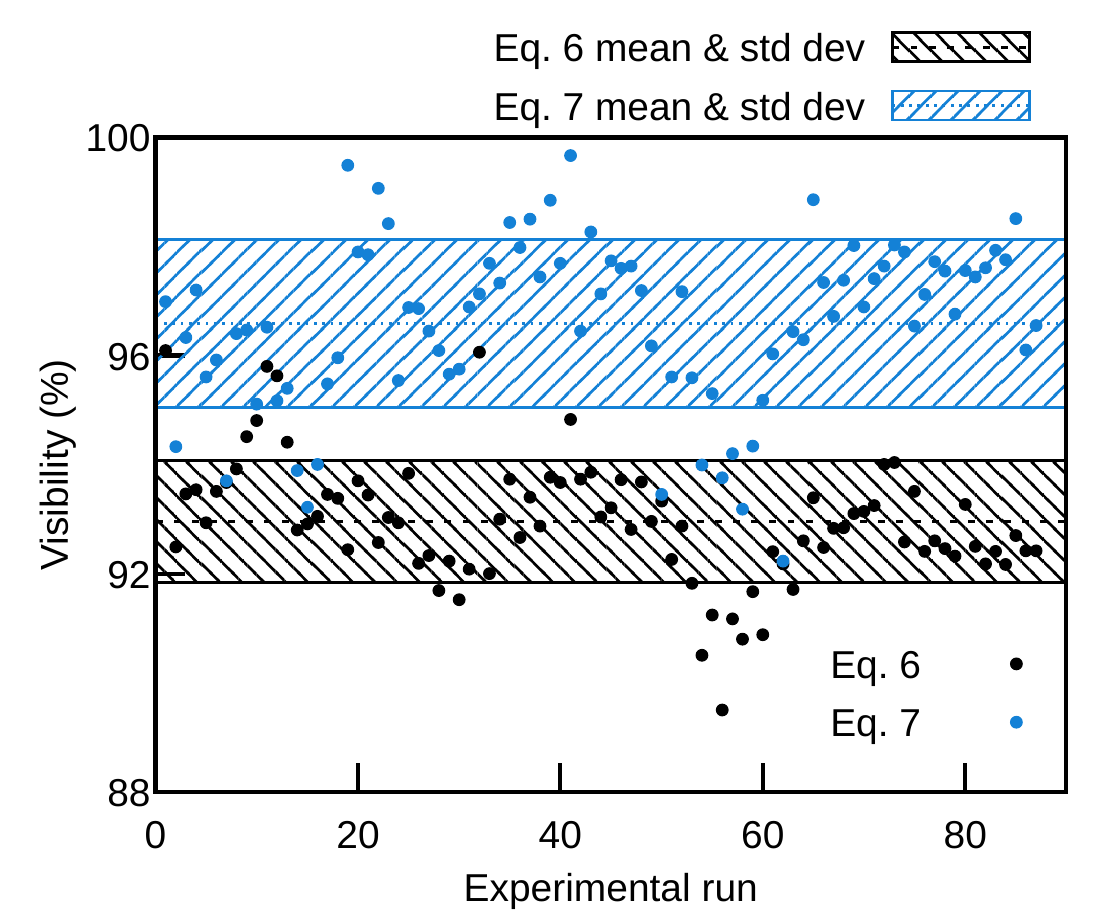}
    }
    \caption{(a) Example coincidence data from the correlated photon source described in~\cite{tang14}, undergoing field testing in near-Space conditions. Raw coincidence data is shown alongside data corrected using both Eq.~\ref{eqn:old-correction-asynch} \&~\ref{eqn:new-correction-asynch}. Visibility (the contrast of the fringes) is seen to increase from 88.8\% to 96.9\%. (b) Updated visibilities plotted alongside the original values taken from~\cite{tang14}. For all experimental runs we observe an increase, from a mean value of 93.0$\pm$1.1\,\% to 96.6$\pm$1.5\,\%. Shaded regions indicate one standard deviation.}
\end{figure}

To assess the impact of this improved correction factor on real-world data, we generated correction factors for a pair of APDs embedded within a correlated photon source operated in a near Space environment~\cite{tang14}. In this particular experiment the detectors were operated in the saturation regime because of the low collection efficiency of the source (discussed in~\cite{chandrasekara15_spie2}). Two fits are plotted using the correction factors calculated from Eqs~\ref{eqn:old-correction-asynch} and~\ref{eqn:new-correction-asynch}, with the visibility of the interference fringes increasing from 88.7\% to 96.9\% (see Fig.~\ref{fig:visibility}). Residual errors are due to the intrinsic extinction ratio of the polarization components within the experimental source. When the full set of experimental data was analysed, this trend of improved visibility was observed over all runs, from a mean value of 93.0$\pm$1.1\,\% to 96.6$\pm$1.5\,\% (see Fig.~\ref{fig:visibilities}).

\section{Concluding remarks}

By considering the proportion of incoming events which encounter a receptive detector, we have developed a model for estimating the accidental coincidences between a pair of asynchronously operated single photon detectors, making use of the concept of an ``effective duty cycle''. By taking this very general approach, we have opened up the possibility of accomodating arbitrary detector recovery processes, as well as arbitrary waiting time statistics of the light source. Unlike approaches which attempt to divide detectors into so-called ``paralyzable'' and ``non-paralyzable'' categories~\cite{schatzel86}, our mechanism is able to capture the full spectrum of partially extending dead-times, as well as any other effects which are included in the model for the detector recovery or the temporal distribution of the incoming events.

We have also shown that by considering only relatively high-level relationships between a few key parameters of our GM-APD circuit, we are able to correct for background coincidences at photon rates commonly considered outside the capabilities of passively quenched systems. The marked improvement in contrast observed in Fig.~\ref{fig:visibility} demonstrates the power of this technique, and the importance of the coincidence noise model when operating at high rates. Once this duty cycle is taken into account and the appropriate accidental coincidences are subtracted, the detectors may be operated at higher rates, effectively increasing their dynamic range.

This treatment can be readily extended to detectors that are actively quenched and to a larger numbers of detectors for multi-photon coincidence applications. Using the numerical approach for calculation, the effective duty cycle can be easily modified for light sources where the photon inter-arrival time has non-Poissonian statistics. The insights gained in this work may aid the deployment of cost-effective and low-power photon counting devices to applications where post-processing of multi-photon events is valid.\\

\section*{Acknowledgements}

This work is supported by the National Research Foundation grant NRF-CRP12-2013-02 \emph{``Space based Quantum Key Distribution''} and the MOE grant MOE2012-T3-1-009 \emph{``Random Numbers from Quantum Processes''}.

\end{document}